\documentclass[aps,psfig,preprint,groupedaddress]{revtex4}
\usepackage{graphicx}
\usepackage{color}

\newcommand{\bee}{\begin{equation}}
\newcommand{\ene}{\end{equation}}
\newcommand{\beea}{\begin{eqnarray}}
\newcommand{\enea}{\end{eqnarray}}
\newcommand{\fpar}[2]{\frac{{\textstyle \partial \/ #1}}{{\textstyle \partial \/ #2}}}
\newcommand{\npar}[3]{\frac{{\textstyle \partial^{#1} \/ #2}}{{\textstyle \partial \/ #3^{#1}}}}
\begin{document}
\title{Stability of nonlinear 1D laser pulse solitons in a plasma}
\author{Vikrant Saxena, Amita Das, Sudip Sengupta, Predhiman Kaw, and Abhijit Sen}
\affiliation{Institute for Plasma Research, Bhat , Gandhinagar - 382428, India }
\date{\today}

\begin{abstract}
In a recent 1D numerical fluid simulation study [Phys. Plasmas {\bf 13}, 032309 (2006)] it
was found that an instability is associated with a special class of  one dimensional nonlinear 
solutions for  modulated light pulses coupled to electron plasma waves in a relativistic cold plasma 
model. It is shown here that the instability can be understood on the basis of the Stimulated Raman 
scattering (SRS) phenomenon and the occurrence of  density bursts in the trailing 
edge of the modulated structures are a manifestation of an explosive instability arising from a 
nonlinear phase mixing mechanism. 

\end{abstract}
 
\maketitle 

\section{Introduction} 
The interaction of intense laser pulses with a plasma has been a topic of research interest 
for decades. The recent advent of ultra high intensity($\sim 10^{20} w/cm^2$) lasers 
has however led to a strong resurgence of this field. The processes of major interest taking place 
during these interactions are 
self focusing, soliton formation, wake-field generation, magnetic field generation
etc. Among those the possibility of coherent nonlinear traveling pulse soliton like 
solutions for such a system has attracted keen attention from physicists both from a 
fundamental research point of view for their possible applications in  diverse areas such  
as particle and photon acceleration, fast ignition concept of laser fusion etc. A large number of 
investigations \cite{kozlov79,ksk92,kuehl93,sudan97,esirkepov98,bulanov,farina01,kala1,vik06} 
have been carried out to study the existence and accessibility of such coherent nonlinear solutions.
In general two classes of soliton solutions have received considerable attention. One having single peak 
in vector potential($R$) as well as in the scalar potential($\phi$) profiles while other having 
multiple peaks of vector potential trapped inside a single peak envelope of scalar 
potential. The single peak one exhibit a continuous spectrum
whereas the one with multiple peaks in $R$ correspond to a discrete spectrum \cite{farina01,kala1}.
For practical applications of these solitonic structures one needs to develop a proper 
understanding of their dynamical properties 
like how they propagate in homogeneous and inhomogeneous plasmas and how they behave if 
subjected to mutual collisions etc.
To reveal the dynamical properties of these 
interesting solutions an attempt 
has been made recently in our earlier work \cite{vik06},  where we dynamically 
evolved these nonlinear solutions with the help of fluid simulations. 

It was found there that the single peak solutions evolve stably in a homogeneous plasma.
They also remain almost unchanged and display nice reflection and transmission properties during 
their course of propagation in an inhomogeneous plasma. 
They were even found to remain intact when subjected to mutual collisions.
On the other hand the solutions with 
multiple peaks in vector potential profile inside a single peak envelope of scalar potential, 
were seen to become unstable after a few tens
of plasma periods by shedding radiation from their trailing edge. Further, these pulses also exhibited
sharp density bursts in their wake region. Such an unstable behavior was also noticed in
 Vlasov simulation\cite{bengt06} of these solutions.

To date the mechanism of this instability and origin of the density bursts remain unclear. 
The present work is devoted to a delineation of the physical mechanism underlying this process. 
We first carry out a detailed study of 
the instability occurring within the pulse extent and demonstrate that the forward Raman 
scattering process is infact responsible for the instability by comparing our simulation 
results with the known analytical values of the growth rates of the Raman forward and backward scattering 
instability. It is seen that the backward scattering growth rates estimated from the simulation
do not match with the corresponding analytical growth rates. Moreover in the simulation 
the wavelengths generated due to scattering processes are larger for the excited 
electrostatic waves than those for the scattered electromagnetic waves which is a clear 
feature of forward Raman scattering. Also backward scattering 
process tends to affect the front edge of the pulse and even in broad pulse it is expected
 to saturate in the leading part of the pulse itself and thus doesn't affect the main body of the 
pulse\cite{sakharov94} which is in contrast to our observations. 
We further provide an understanding of the density bursts observed in the wake of 
the pulse by employing a simple model calculation based on the relativistic wave breaking 
phenomenon. In particular, we try to predict the approximate time between two consecutive 
bursts on the basis of the calculation of the mixing time for two coexisting relativistic 
plasma waves and then compare it with the corresponding time observed in our simulations.
 A detailed investigation in this regard is presented.  

In the next section a 1D model for the interaction of a relativistically intense 
laser pulse with a cold collisionless plasma with fixed ion background is described. 
In the same section we also recapitulate the possible solutions briefly.  Then 
in the following section, a 
detailed analysis of the instability is provided. Further in section - IV we provide an understanding of 
the density bursts that are observed in the wake of the moving multipeak solution on the basis of the
relativistic wave breaking phenomenon. Finally in the last section we present the conclusions 
of the paper.

\section{Basic Equations and Stationary Solutions}

The basic equations are the relativistic set of fluid evolution 
equations for a cold plasma in one dimension together with the Maxwell equations for 
the electromagnetic wave. We consider spatial variations to exist only 
along $x$, the direction of propagation, and consider the ions to be stationary. The relevant 
set of fluid and field equations are then, 

\bee
\fpar{n}{t} + \fpar{ (nu)}{x} = 0.
\label{d_cont} 
\ene
\bee
    (\fpar{}{t} + u \fpar{}{x})(\gamma u) = \fpar{\phi}{x} - \frac{1}{2 \gamma} \fpar{A_{\perp}^{2}}{x}
\label{l_mom} 
\ene
\bee
\npar{2}{\phi}{x} = n - n_0(x)
\label{pois}
\ene
\bee
   \npar{2}{\vec{A}_{\perp}}{x} - \npar{2}{\vec{A}_{\perp}}{t} =  \frac{n\vec{A}_{\perp}}{\gamma} 
\label{wave}
\ene

where (\ref{d_cont}) is the electron continuity equation, (\ref{l_mom}) is the parallel electron 
momentum equation,
(\ref{pois}) is the Poisson's equation for the electrostatic potential $\phi$, (\ref{wave}) is 
the wave equation
for the vector potential $\vec{A}_{\perp}$ and other notations are standard. 
The perpendicular electron momentum equation has been integrated exactly to obtain the 
conservation of the transverse 
canonical momenta (sum of particle and the field momenta) as 
$u_{\perp} - \vec{A}_{\perp}/\gamma = 0$ and used to
eliminate $u_{\perp}$ in the above equations. 
Here $\gamma$ is the relativistic factor 
$$
\gamma = \sqrt{\frac{1 + A_{\perp}^{2}}{1 - u^{2}}}
$$
 In writing the above equations we have chosen to normalize 
the density by some appropriate density $n_{00}$. The length is normalized by the 
corresponding skin depth 
 $c/\omega_{pe0}$ (where $\omega_{pe0} = \sqrt{4\pi n_{00}e^2/m_e}$) and time by the 
inverse of the plasma frequency 
$\omega_{pe0}^{-1}$. The scalar and vector potentials are normalized by $mc^2/e$. 
In Poisson's equation 
$n_0(x)$ corresponds to the background ion density normalized by $n_{00}$.

The coupled set of nonlinear equations(\ref{d_cont}- \ref{wave}) 
 permit a variety of coherent solutions. A class of one dimensional propagating solutions 
with modulated envelope structure 
of the above set have been obtained in the past by using  the coordinate transformation $\xi = x - \beta t$ 
 and $\tau = t$ (where $\beta$ represents the group velocity of the structure). The vector 
 potential is assumed to be circularly polarized and has a sinusoidal phase variation of the 
form 
 $ \vec{A} = (a(\xi)/2)[\{\hat{y} + i\hat{z} \}\exp(-i \lambda \tau ) + c.c. ]$. The plasma
oscillations associated with the envelope structure are assumed to have no dependence on $\tau$.
 This is the so called electrostatic approximation which is valid if there is not a significant change 
 in the plasma parameters within the pulse duration. 
The above transformations convert Eqs.(\ref{d_cont},\ref{l_mom}) into ordinary differential 
equations 
which can be integrated to give $n(\beta - u) = \beta$ and $\gamma(1-\beta u) - \phi = 1$, where 
one assumes that at the boundaries $u = 0$, $\phi = 0$ and $n = 1$. One eliminates $n$ to write 
Poisson's equation [see Eq.(\ref{pois})] as 
\bee
\phi^{\prime \prime} = \frac{u}{(\beta -u)}
\label{pois_st}
\ene
Here prime($^{\prime}$) denotes derivative with respect to $\xi$. 
Writing $a(\xi) = R exp(i \theta)$, the wave equation [see Eq.(\ref{wave})] can be written as 
\bee
R^{\prime \prime} + \frac{R}{1-\beta^2} \left[\left(\lambda^2 -\frac{M^2}{R^4}\right) \frac{1}{1-\beta^2} 
- \frac{\beta}{\beta - u} \frac{1-\beta u}{1+\phi}\right] = 0
\label{wave_st}
\ene
Here $M = R^2[(1-\beta^2)\theta^{\prime}-\lambda \beta]$ is a constant of integration and 
$R^2 = A_y^2 +A_z^2$. 
Eqs.(\ref{pois_st},\ref{wave_st}) form a coupled set of second order differential equations 
in two fields $\phi$ and $R$ respectively. The longitudinal velocity $u$ appearing in the two 
equations 
can  be expressed entirely in terms of $R$ and $\phi$ as 
\bee
u = \frac{\beta(1+R^2) - (1+\phi)[(1+\phi)^2 - (1-\beta^2)(1+R^2)]^{1/2}}{(1+\phi)^2 + \beta^2(1+R^2)}
\label{u_st}
\ene
Eqs.(\ref{pois_st},\ref{wave_st}) have been solved by Kaw {\it et al.} \cite{ksk92} and others 
\cite{kala1} for $M = 0$. 

In the absence of any further simplifying assumptions, above equations [see Eqs.(\ref{pois_st},\ref{wave_st})]
cannot be solved analytically. However, for the general case several varieties of 
numerical solutions have been obtained(for M = 0). A detailed characterization of some of 
these solutions on the basis of group speed $\beta$ and the frequency parameter 
$\lambda$ has been made in some of the earlier studies \cite{ksk92,farina01,kala1} 
where $\lambda$ is defined as $\omega(1 - \beta^2)$.

We reproduce the $\lambda-\beta$ spectrum and show the two varieties of the solutions with 
that in Fig.1. 
A continuum spectrum  in the $\lambda - \beta$ plane has been observed only for those 
solutions which have a single peak of vector potential $A$ as well as electrostatic potential 
$\phi $. These solutions have a  reasonably lower amplitude and satisfy $\phi < A$. 
On the other hand there are solutions which occur only for discrete values of $\lambda$ 
for a given $\beta$. These solutions differ from those with continuum spectrum as they 
have several multiple peaks of the vector potential $A$ trapped inside an envelope 
of $\phi$. The scalar normalized potential  $\phi >> A$ for these solutions. 
The electron density in the central region is  strongly evacuated and the light wave is 
trapped in this density cavity.

In our recent numerical work\cite{vik06} it was found that the solutions corresponding 
to the continuous spectrum and having single peak in vector potential display robustness 
during their propagation in a homogeneous as well as in an inhomogeneous plasma while those 
admitting a discrete spectrum and having multiple peaks in the vector potential
tend to develop an instability in their trailing edge even while propagating in a homogeneous 
background plasma. In the next section we carry out 
a detailed investigation of this characteristic instability 
of the multipeak solutions and compare our simulation results, in particular the growth rate 
of the instability, with the corresponding analytical results 
for forward stimulated Raman scattering(fSRS) instability.

\section{Characteristic instability of the Multipeak Solutions : A detailed investigation}

 As mentioned above, the structures with multiple peaks in $R$ exhibit an interesting 
instability wherein the perturbed fields are 
ejected from the trailing edge of the solutions as shown in Fig.2. 
In various subplots of the figure, profiles of
vector potential as well as scalar potential are shown at different time instants.
It is obvious from the figure that the structure seems to evolve stably for few tens of 
plasma periods before it starts emitting from the trailing edge.

In Fig.3 we show the growth of the perturbations at three different time instants, viz. 
$t = 100, 110$, and $120$ electron plasma periods, measured as the difference between exact and the numerically 
observed value of the scalar potential $\phi$ and the vector potential $R$ within the 
structure. The structure has been identified by 
simultaneously plotting the equilibrium electron density curve translated by $\beta t$
(the  cavitation in electron density essentially 
provides the spatial extent of the structure). The pulse is moving 
towards right with a group speed of $\beta=0.8$.It can be observed from the figure that a 
small amplitude perturbation starts at the  front edge of the pulse. It  suffers  
continuous  amplification as it trails behind towards  the rear edge. 
Finally from the rear edge of the solutions, structures get 
ejected and this process continues. \\


Let us first have some understanding of the  mechanism for this instability 
on the basis of the well known relativistic stimulated Raman 
scattering (SRS) phenomena. The physical mechanism of  SRS is simple 
and can be understood by realizing that an incident electromagnetic  
radiation  generates  a scattered
light wave due to the transverse currents in the plasma medium. The nonlinear interaction 
of the scattered light wave with the incident light pulse in turn produces an electrostatic 
plasma wave. The plasma wave can get resonantly excited to a very large amplitude if 
appropriate frequency matching conditions are satisfied. The instability can only get 
excited provided 
a threshold condition on the electron density is satisfied which arises from the condition 
$\omega \ge \omega_{pe}/\sqrt(\gamma)$. Here $\omega$ and $\omega_{pe}$ are the laser 
frequency and the 
electron plasma wave frequency respectively and  $\gamma $ is the relativistic factor. 
This shows that the instability can only be excited provided the electron density satisfies 
the condition of  $n_e \le n_{th}$, where $n_{th} = \gamma \omega^2/4$ within a reasonable 
spatial extent to observe several e - foldings in the growth.\\

Before we provide a comparison for the analytical growth rates 
of forward and backward Raman scattering with the numerically observed growth rates, 
let us discuss the well established theoretical results for the two cases. 
The growth rate  for the relativistic Raman forward 
scattering instability for which the scattered  wave moves in the same direction 
as the incident light pulse is given by the following expression 
\cite{mori94,sakharov94,guerin95,decker_pop96},\\

\begin{equation}
\Gamma_{rfs} = \frac{1}{2\sqrt{2}\omega}\frac{A_0}{(1+A_0^2/2)} \\
\end{equation}
On the other hand the growth rate of the backward Raman instability
for which the scattered wave moves opposite to the incident pulse is given by
two different expressions in two different regimes\cite{sixauthor,decker_pop96}. When the 
 condition $v_{osc}/c < (\omega_{pe}/\omega)^{1/2}$
 which in relativistic case becomes $A_0^2/\gamma^{3/2} < \omega_{pe}/\omega$, is satisfied 
the growth rate for the backward Raman scattering instability is given by 
\begin{equation}
\Gamma_{brs} = \frac {\sqrt{\omega}}{4} \frac{A_0}{(1+A_0^2)^{5/8}}
\label{brs1}
\end{equation}
and when $A_0^2/\gamma^{3/2} > \omega_{pe}/\omega$ holds the expression for the growth rate 
for the backward Raman scattering instability reads
\begin{equation}
\Gamma_{brs} = \sqrt{3} \left(\frac{\omega}{16}\right)^{1/3} \frac{A_0^{2/3}}{(1+A_0^2)^{1/2}}
\label{brs2}
\end{equation}
Here, $\omega$ is the frequency of the light pulse and $A_0$ is the maximum amplitude of 
the vector potential. We compare the analytical growth rates for 
both kinds of the Raman scattering instability with those 
evaluated from the results of the  numerical simulations in Fig.4. 

The growth rate from the observed data is calculated using the expression
for the amplification factor for the parametric instability 
\cite{nishikawa_app}
\bee \alpha = exp(\Gamma_{sim} L / (V_1 V_2)^{1/2}) \ene 
where $\Gamma$ is the growth rate of the parametric instability, L is the
length of the interaction region, and $V_1$,$V_2$ are the relative group
speed of the daughter waves measured with respet to the pump wave. In our 
simulations we observe both the daughter waves almost standing together
behind the pump which leads to $V_1 = V_2 = \beta$ and the expression for the 
amplification factor reduces to
\bee \alpha = exp(\Gamma_{sim} L / \beta) \ene 
The amplification factor is the ratio of the final to the initial 
perturbation amplitudes in scalar potential $\phi$.\\

Now for a comparison, numerical growth rates from the simulations and analytical growth rates for 
the two cases viz. the forward and backward Raman scattering instability are obtained for $8$ 
different solutions as discussed above. These solutions differ from each other 
with respect to the number of light wave 
peaks associated with them, the peak vector potential amplitude and its  
frequency. These  parameter details  of various  solutions have been presented in Table - I
together with the analytical growth rates for the forward as well as for the backward SRS 
and the growth rates obtained from the simulations. 
The table also shows the value of the threshold density $n_{th}$ and the minimum electron 
density 
$n_{min}$ for the solutions. Note that for all the cases $n_{min} $ is less than $n_{th}$.
so that the threshold criterion for the excitation of SRS is adequately satisfied. 
The upper subplot of 
Fig.4 shows a plot of $\Gamma \omega$ with $A_0$ for the theoretical growth rates for
forward Raman scattering instability as well as numerical growth rates. 
It is clear  from the figure that for all  varieties of multipeak solutions, the growth rate 
of the observed instability agrees closely  with  the analytical value of the forward SRS. 
In the lower subplot of Fig.4 we compare the theoretical growth rates for backward Raman 
scattering instability with the numerically observed growth rates. 
In fact we plot $\Gamma/ \sqrt{\omega}$ with $A_0$, corresponding to $8$ multipeak solutions 
for the theoretical growth rates 
for backward Raman scattering instability as well as for the 
numerical growth rates. We observe that there is a clear mismatch in the values.\\ \\

\vspace*{-0.1in}
\begin{table}[h]
\caption{Comparison of instability's growth rates observed in simulations
with the theoretical values. Here $\beta,\omega,p,A_0,w,n_{min},n_{th},\Gamma_{frs}, 
\Gamma_{brs}$ and $\Gamma_{sim} $ respectively stand for group speed and frequency of the pulse, 
number of extrema and peak amplitude of the vector potential, transition width for growth, 
minimum electron density value in the 
cavity, threshold electron density for Raman scattering, theoretically estimated growth rates 
for forward and backward SRS and numerically observed growth rates of the instability. }
\vspace*{0.2cm}
\begin{tabular}{ c  c  c  c  c  c  c  c  c  c }
\hline
\hline
\hspace*{0.3cm}$\beta$ \hspace*{0.3cm}&\hspace*{0.3cm} $\omega$ \hspace*{0.3cm}&\hspace*{0.3cm} $p$ \hspace*{0.3cm}&\hspace*{0.3cm} $A_0$ \hspace*{0.3cm}&\hspace*{0.3cm} $w$ \hspace*{0.3cm}&\hspace*{0.3cm} $n_{min}$ \hspace*{0.3cm}&\hspace*{0.3cm} $n_{th}$ \hspace*{0.3cm}&\hspace*{0.3cm} $\Gamma_{frs}$ \hspace*{0.3cm}&\hspace*{0.3cm} $\Gamma_{brs}$ \hspace*{0.3cm}&\hspace*{0.3cm} $\Gamma_{sim}$  \\
\hline
$0.9$ & $2.14266$ & $3$ & $1.5343$ & $6$ & $0.5636$ & $1.6053$ & $0.1163$ & $0.6437$ & $0.1547$ \\
$0.8$ &  $1.4803$ & $3$ & $2.2522$ & $7$ & $0.4762$ & $1.1559$ & $0.1521$ & $0.5462$ & $0.1597$ \\
$0.8$ &  $1.4371$ & $4$ & $3.1362$ & $9$ & $0.4539$ & $1.4974$ & $0.1304$ & $0.5049$ & $0.1752$ \\
$0.8$ & $1.403$ & $5$ & $4.046$ & $12$ & $0.4488$ & $1.8501$ & $0.111$ & $0.4688$ & $0.1532$  \\
$0.6$ &  $0.93522$ & $5$ & $7.2710$ & $17$ & $0.3756$ & $2.0927$ & $0.1002$ & $0.3437$ & $0.1346$ \\
$0.4$ &  $0.6943$ & $4$ & $10.3772$ & $19$ & $0.2859$ & $1.9353$ & $0.0964$ & $0.2778$ & $0.0609$ \\
$0.5$ &  $0.67225$ & $7$ & $19.1134$ & $35$ & $0.3334$ & $3.410$ & $0.05473$ & $0.2249$ & $0.0255$ \\
$0.5$ &  $0.59288$ & $9$ & $31.4606$ & $50$ & $0.3333$ & $3.1188$ & $0.037833$ & $0.1828$ & $0.00852$ \\
\hline
\hline
\end{tabular}
\end{table}
\vspace*{0.2cm}

We note from the figure and from the table as well, that for forward SRS 
instability the theoretical values of the growth rates match well with the 
numerically observed values which is not the case for the backward SRS 
instability. It should also be noted that 
the scattered light wave structures are found to remain trapped
inside the plasma wave structures which trail behind with respect to the moving pulse and 
the scale length of the scattered light wave amplitude $dR$ is shorter than that of the 
perturbed scalar potential $d \phi$ (as is clear from Fig.3) which are clear signatures 
of forward Raman scattering instability. Also the backward Raman instability is known to 
get saturated in the leading edge of the pulse itself thus not affecting the main body of 
the pulse\cite{sakharov94} which is in contrast with our observations. Therefore we 
 discard the possibility of the backward Raman scattering to be a potential candidate for
the depletion of the pulse. It should be noted here that even though the growth rate of the 
backward Raman scattered instability is more than the forward SRS the solutions seem to exhibit 
only the forward SRS instability.\\

To provide an additional evidence of the forward Raman scattering instability 
we display in Fig.5 the frequency spectrum of one component of the 
vector potential ($A_y$) measured both in the laboratory frame as well as in the frame moving 
with the pulse. As evident from the left subplot of Fig.5, the spectrum measured in the 
laboratory frame has peaks at pump wave frequency
[$\omega_0 = \lambda/(1-\beta^2) $], plasma frequency($\omega_{pe}$), sideband frequencies
($\omega_0 - \omega_{pe}$ and $\omega_0 + \omega_{pe}$ ) and at another frequency 
$2\omega_0 - \omega_{pe}$ which might result from the interaction between the left sideband 
and the pump wave frequency. The spectrum also affirms the presence of forward Raman
scattering as we obtain both the frequency sidebands. It also supports our conclusion
 discarding the possibility of the backward Raman scattering as in the backward Raman
 scattering the power in the up-shifted frequency band is usually negligible\cite{kruer} 
 which is in contrast with the present frequency spectrum.
In the right subplot of the same figure where the spectrum
measured in the moving frame is shown, there is only one peak at a frequency 
$\omega_0 - k\beta$. The result is as per our expectations because while measuring the 
frequencies in the moving frame of the pulse the Doppler effects come into picture.
 The reason why we don't observe other Doppler shifted frequencies is also quite simple 
 since if we fix a position with respect to the head of the pulse of the observation point 
 we can only observe the frequency with which the wave 
 vector components are oscillating ($\omega_0$ in the present scenario).\\

We now try to understand the total absence
of the forward SRS instability for the single peak 
variety of the modulated solutions. For all the single peak  solutions the scalar electrostatic 
potential $\varphi$ is observed to be very weak in comparison to the vector potential $a$. 
The smaller  value of $\varphi$ 
for these solutions is due to the fact that  electron cavitation in these structures are much 
weaker in comparison to the multipeak solutions. Table II provides a detailed description of the 
parameters of this variety of solutions. Note that even though the forward SRS growth rate of the 
solutions obtained from the analytical expression may be finite, but in all cases the threshold condition on density for the excitation of forward 
SRS is not satisfied and the growth rate has no meaning. This is the primary reason for 
the robustness of these structures in fluid simulations.  Moreover, in the small 
amplitude limit these solutions are similar to the exact nonlinear Schrodinger 
soliton solutions which are known to be stable. \\ \\
 
\begin{table}[h]
\caption{Comparison of minimum electron density and threshold density for RFS for single
peak solitons}
\vspace*{0.2cm}
\begin{tabular}{ c c c c c }
\hline
\hline
\hspace*{0.3cm}$\beta$ \hspace*{0.3cm}&\hspace*{0.3cm} $\omega$ \hspace*{0.3cm}&\hspace*{0.3cm} $A_0$ \hspace*{0.3cm}&\hspace*{0.3cm} $n_{min}$ \hspace*{0.3cm}&\hspace*{0.3cm} $n_{th}$ \\
\hline
$0.1$ & $0.9495$ & $0.7350$ & $0.9392$ & $0.2797$ \\
$0.2$ & $0.9896$ & $0.5209$ & $0.9822$ & $0.2720$ \\
$0.3$ & $1.0329$ & $0.3509$ & $0.9959$ & $0.2827$\\
$0.4$ & $1.0857$ & $0.2002$ & $0.9995$ & $0.3005$\\
$0.5$ & $1.1453$ & $0.2585$ & $0.9984$ & $0.3387$ \\
$0.8$ & $1.0066$ & $0.1158$ & $0.9998$ & $0.6966$ \\
\hline
\hline
\end{tabular}
\end{table}

In the next section, we present a model calculation based on phase mixing/wave breaking phenomenon to 
explain the appearance of bursts in the wake of the moving multiple peak solution.

\section{A model calculation for the density Bursts in the wake of the multipeak solution}

As a result of forward Raman instability the small perturbations in the front end of 
the pulse are continuously amplified before being ejected from the rear end. We observe
that following this process there appear density spikes in the wake of the moving pulse. 

A plausible explanation for the appearance of density spikes lie in the basic nonlinear
effect associated with a relativistically intense plasma oscillation, where due to relativistic
variation of electron mass, the effective plasma frequency becomes a function of position \cite{infeld89}.
As a result different fluid elements which are
at different locations in space oscillate at different frequencies leading to loss of coherence due 
to phase mixing. Ultimately there occurs 
a point in time when two adjacent oscillating fluid elements cross through each other and 
the initial coherent oscillation explosively breaks ( wave breaking ). 
This
effect is very similar to the phase mixing of non-relativistic 
plasma oscillations as discussed
by Dawson et. al. \cite{dawson59} for an inhomogeneous plasma, where the
spatial dependence of plasma frequency arises due to background inhomogeneity.

In the present case however, instead of a plasma oscillation, a spectrum of intense plasma waves are 
excited in the wake of the multipeak solution. Based on the above physics, we present below a 
model calculation where we treat the evolution of two relativistically intense plasma waves whose wave 
numbers differ by an amount $\Delta k$. This is an extension of Infeld's \cite{infeld89} calculation for a
relativistically intense plasma oscillation. In order to represent the scenario behind the moving pulse at
a time when the plasma waves are ejected out, we take the initial density and velocity perturbations as

\bee
\delta n_{e}(x,0) \approx \Delta \cos(\frac{\Delta k x}{2}) \cos((k + \frac{\Delta k}{2})x)
\ene
and 
\bee
v_{e}(x,0) \approx \frac{\omega_{pe} \Delta}{k} \cos(\frac{\Delta k x}{2}) \cos((k + \frac{\Delta k}{2})x)
\ene

We now study the evolution of these perturbations. In Lagrange coordinate, the relativistic equation of 
motion of a fluid element is given by \cite{kruer90}.
\bee
\frac{\ddot{\xi}}{(1 - \dot{\xi}^{2}/c^{2})^{3/2}} + \omega_{pe}^{2} = 0
\ene
where $x = x_{0} + \xi(x_{0},\tau)$, $\xi$ being the displacement of a fluid element from its 
equilibrium position ($x_{0}$). Using (13) and (14) as initial condition, in the weakly relativistic
limit ($\omega_{pe} \Delta / c k \ll 1$) and ($\Delta k / k \ll 1$), solution of equation (15) may be 
written as
\bee
\xi(x_{l},\tau) \approx \frac{\Delta}{k} \cos(\frac{\Delta}{2} x_{l}) \sin(\tilde{\omega}_{pe} \tau - 
k x_{l})
\ene
where $x_{l} = x_{0} + \xi(x_{0},0)$ is a new Lagrange coordinate and
\bee
\tilde{\omega}_{pe} \approx \omega_{pe} \left[ 1 - \frac{3}{16} \frac{\omega_{pe}^{2}
\Delta^{2}}{k^{2} c^{2}} \cos^{2}(\frac{\Delta k}{2} x_{l})\right]
\ene
Dependence of $\tilde{\omega}_{pe}$ on the initial position $x_{l}$ is a clear indication of phase mixing.
Using Poisson's equation, the electron density $n_{e}$ can be expressed in terms of $\xi(x_{l}, \tau)$ as
\bee
n_{e}(x_{l},\tau) = \frac{n_{0}}{1 + \frac{\partial \xi / \partial x_{l}}{1 - \partial \xi / 
\partial x_{l}|_{\tau=0}}}
\ene
Substituting the expression for $\xi(x_{l},\tau)$ in the above expression, the electron density in terms of
new Lagrange coordinate $(x_{l}, \tau)$ finally stands as 
\bee
n_{e}(x_{l},\tau) \approx 
\frac{n_{0} \{1 + \Delta \cos(\frac{\Delta k}{2}x_{l})\cos kx_{l}\}}
{1 + \Delta \cos(\frac{\Delta k}{2}x_{l})[\cos kx_{l} + \cos(\tilde{\omega}_{pe} \tau - kx_{l})
\{\frac{\tau}{k}\frac{\partial \tilde{\omega}_{pe}}{\partial x_{l}} - 1\}]}
\ene
The presence of secular term in the denominator clearly shows that the electron density
will eventually explode in a time scale 
$\omega_{pe} \tau_{mix} \sim \left(\frac{3\omega_{pe}^{2} \Delta^{3}}{16 k^{2} c^{2}} (\frac{\Delta k}{k})\right)^{-1}$. 
This time scale 
depends on the level of density fluctuation $\Delta$ and the spread ``$\Delta k$'' of the
plasma waves.  We would like to emphasize here, that in this case wave 
breaking happens at arbitrarily low amplitudes ($\Delta$). This is in contrast
to the earlier works on wave breaking \cite{akhiezer56,coffey71,modena02}, 
which require the wave to reach a critical amplitude before it breaks.

We now apply the above expression for phase mixing time to estimate the time between consecutive density 
bursts and compare it with our numerical solution. We calculate the values of $k$ and $\Delta k$ from the 
$k-$spectrum of the density profile of a relaxed state in between two bursts.
Fig.7 shows the frequency spectrum of such a relaxed state ( no bursts) at $t = 112.1 \omega_{pe}^{-1}$.
In this state the maximum amplitude of the plasma waves is $\Delta \sim 4.0$ and from their k-spectrum 
 $k \sim 1.3 \omega_{pe}/c$ and  $\Delta k \sim 0.2 \omega_{pe}/c$. 
This gives $\omega_{pe} \tau_{mix} \sim 1$. 
This implies that a burst should be observed at $\omega_{pe}t \sim 113.1$ and indeed such a burst is 
observed as shown in Fig.6 where two consecutive 
density bursts occurring at times $t = 110.9 \omega_{pe}^{-1}$ and $t = 113.5 \omega_{pe}^{-1}$ 
are shown together with the in between relaxed state at $t = 112.1 \omega_{pe}^{-1}$(dotted curve).
If we assume that the relaxation time for a density spike to be of the same order as the time required for 
the formation of a spike from a relaxed state, 
then the time between two consecutive bursts turns out to be 
$\sim 2  \omega_{pe}^{-1}$ which is in close agreement with the temporal spacing observed between two 
consecutive bursts (Fig.6). The size of density spike observed in our
simulation is limited by the grid size $\Delta x$ as $\delta n / n_{0} \sim 1/
\Delta x$. In the limit $\Delta x \rightarrow 0 $, $\delta n / n_{0} 
\rightarrow \infty$ as is really the case with wave breaking of a plasma wave
in a cold plasma.

\section{Summary}
In the present paper we have investigated the instability responsible for the 
break up of the multihump solution and we identify it to be the forward Raman 
scattering instability. The growth rate obtained from the simulations is compared with
the theoretically estimated growth rates for both the forward and the backward Raman 
scattering instabilities and found to match well with those for the forward SRS. 
We also present and explain the Fourier spectrum of the scattered electromagnetic fields 
which also supports our arguments about the forward Raman scattering.
Furthermore, we provide an explanation for the density bursts observed in the wake of 
the moving multipeak solutions by means of a model calculation based on the relativistic 
wave breaking phenomenon.\\

\section{Acknowledgment}
We would like to acknowledge the valuable discussions with W. L. Kruer
and P. N. Guzdar. This work was financially supported by DAE-BRNS sanction 
no.: 2005/21/7-BRNS/2454.


\newpage

\begin{center}
{\bf FIGURE CAPTIONS}
\end{center}

\noindent
FIG.1 : The $\lambda - \beta$ spectrum (in subplot 'S') and two possible variety of solutions. The plot in 
subplot tagged with 'A' is for $\beta = 0.05, \lambda = 0.92$ and corresponds to the continuous
spectrum in the subplot 'S'. The other subplot tagged with 'B' is for $\beta = 0.8, 
\lambda = 0.50518049$ and corresponds to the discrete spectrum, in particular the dotted line with 'stars' .\\

\noindent
FIG.2 : The scalar potential, vector potential and electron density profiles
 of a multipeak solution with $\beta = 0.8, \lambda = 0.50518049$ are shown at 
 four different times viz. $t = 0, 80, 100, 120$ electron plasma periods.
 The development of the instability in the trailing edge as well as the 
 consequent density bursts are evident in the two lower subplots.\\

\noindent
FIG.3 : The growth of the perturbation in the scalar potential amplitude 
$\phi$ (solid curve in subplots of the left column) and in the vector 
potential amplitude $R$  (solid curve in subplots of the right column) 
is shown respectively in left and right column of subplots at three different
times viz. t = 100, 110 and 120 electron plasma periods . It is clear
that there appear smaller wavelengths(smaller k's) in the scattered radiation
than in the excited electrostatic oscillations which is a clear signature 
of forward Raman instability. Also shown is the density cavity associated 
with the original pulse translated with $\beta$  (dotted curve in all the 
subplots of this figure).\\

\noindent
FIG.4 : Comparison of the growth rates estimated from the simulation(triangles) 
with the analytical value of the growth rates(circles) of the relativistic forward 
Raman instability(upper subplot) and the backward Raman instability(lower subplot) for $8$ different solutions.
The simulation values match well with the analytical values for forward Raman scattering instability and they don't match with the analytical values for 
backward Raman scattering instability.\\

\noindent
FIG.5 : The two subplot show the Fourier power spectrum in frequency. The 
left subplot correspond to the lab frame whereas the right one is for the 
pulse frame. In the lab frame five we get peaks at $\omega_0 - \omega_{pe}$,
$\omega_{pe}$,$\omega_0$,$2\omega_0 - \omega_{pe}$ and $\omega_0 + \omega_{pe}$.
On the other hand in the pulse frame the spectrum comprises of a single peak at 
a Doppler shifted frequency $\omega_0 - k\beta$.\\

\noindent
FIG.6 : Two consecutive density bursts in the wake of the pulse at t = 110.9 and 113.5 electron
plasma periods together with the in between relaxed state at t = 112.1 electron plasma periods (dotted line).
\\

\noindent
FIG.7 : An expanded view of the k-spectrum of the relaxed density at $t = 112.1$ in 
between two consecutive density bursts shown in previous figure. We note that the dominant k has the 
value $\approx 1.3$ and the full width at half maximum gives $\Delta k \approx 0.2$.\\

\newpage
\begin{figure}
\includegraphics[]{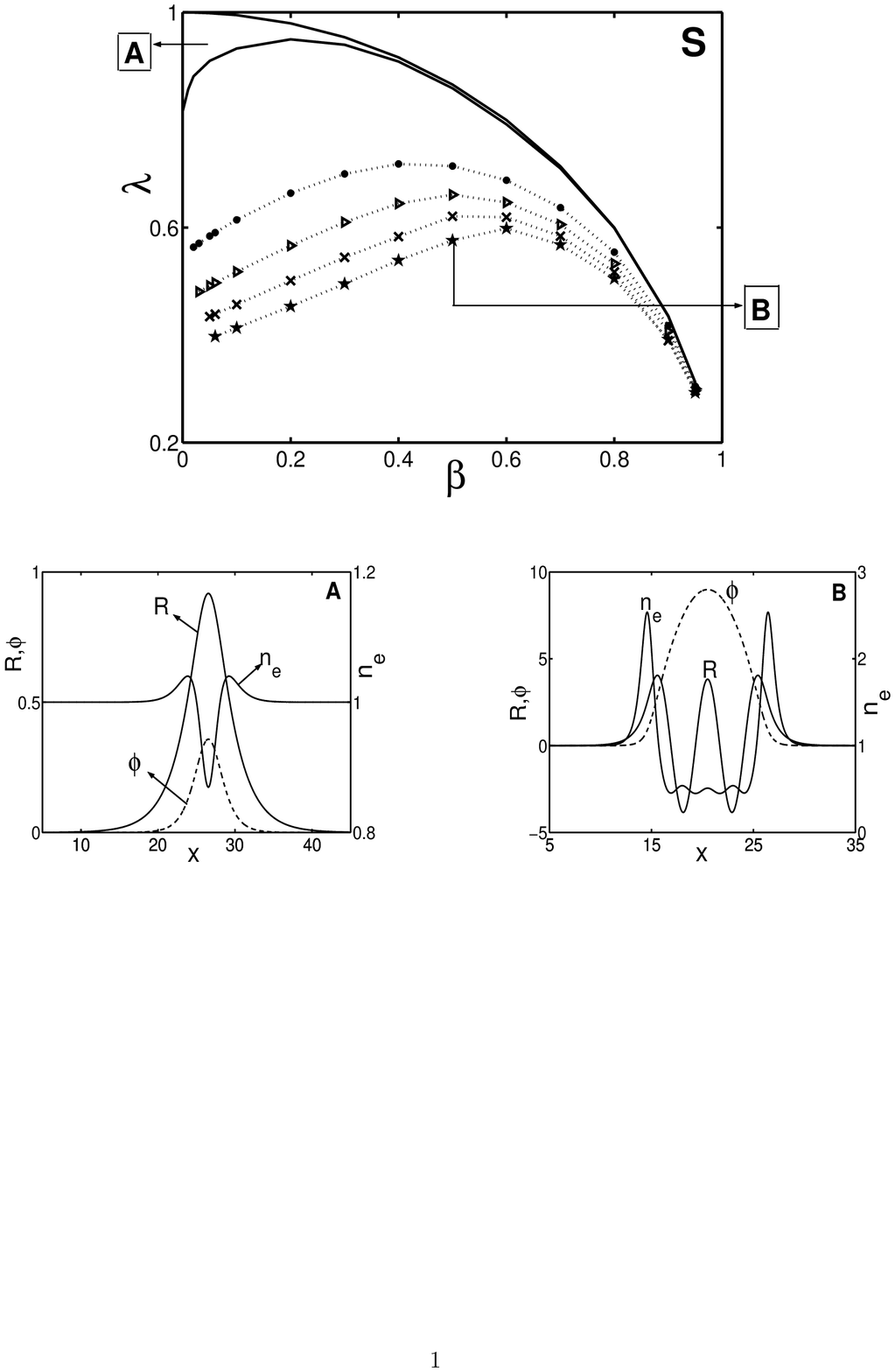}
\end{figure}

\newpage
\begin{figure}
\includegraphics[]{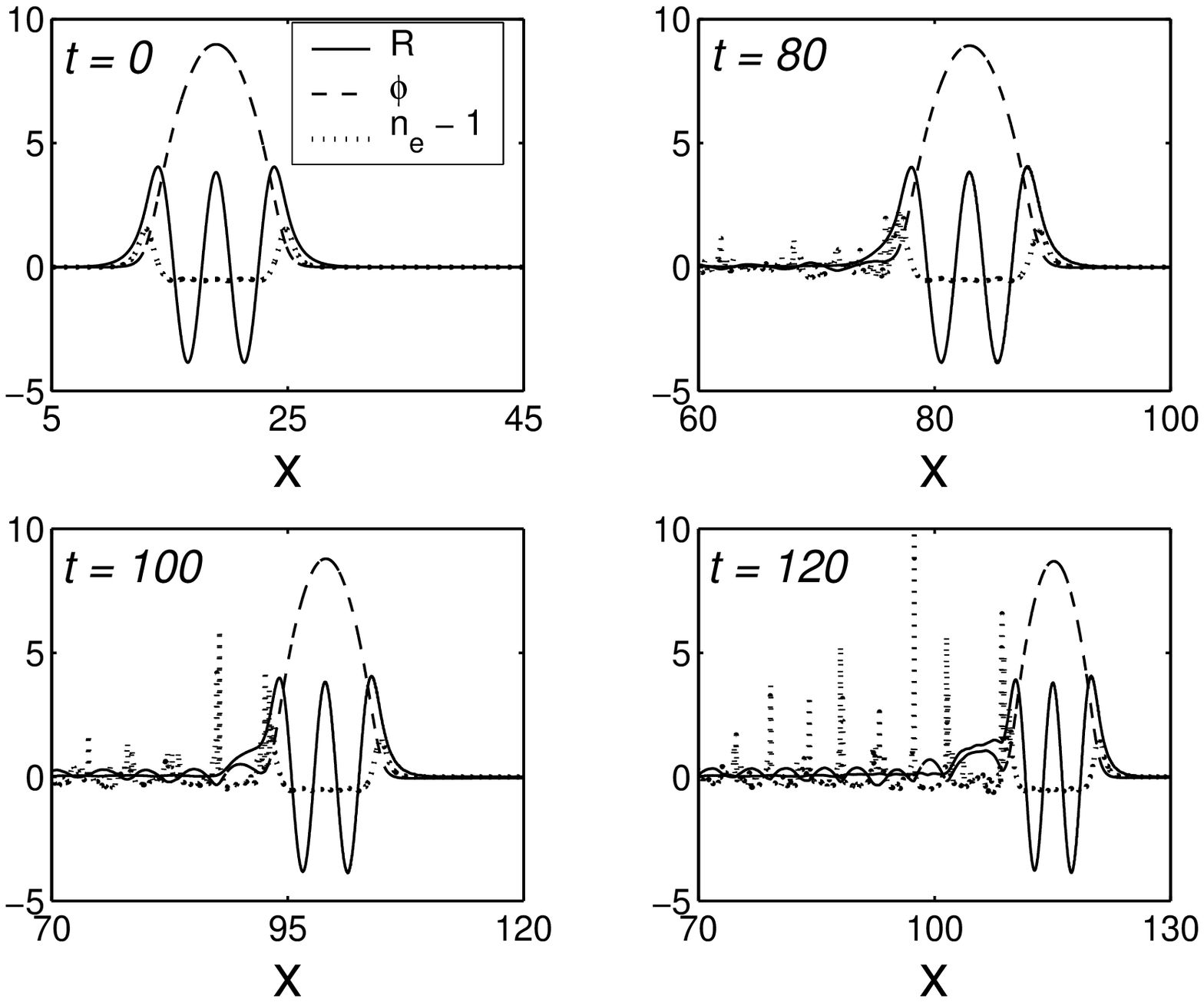}
\end{figure}

\newpage
\begin{figure}
\includegraphics[]{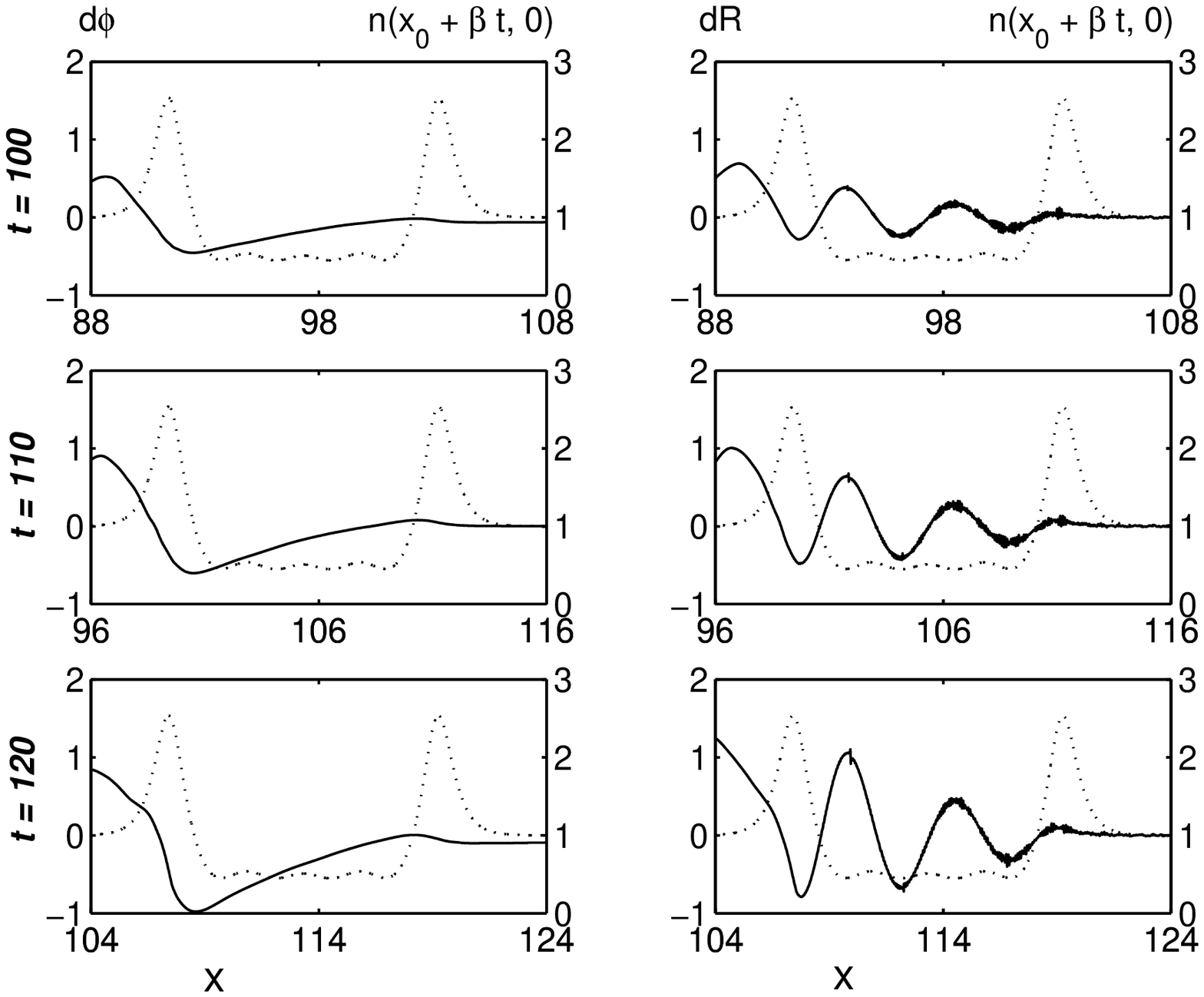}
\end{figure}

\newpage
\begin{figure}
\includegraphics[]{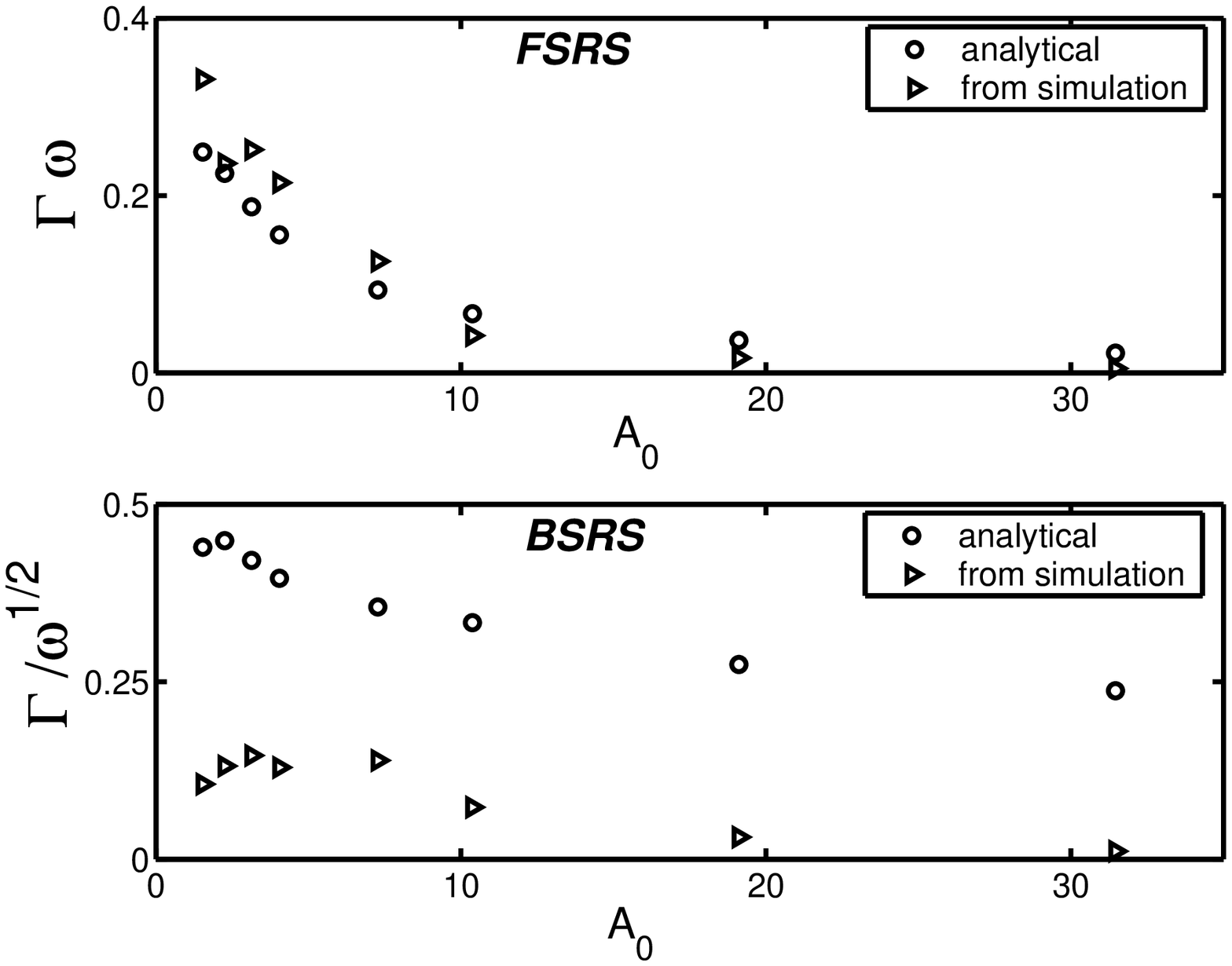}
\end{figure}

\newpage
\begin{figure}
\includegraphics[]{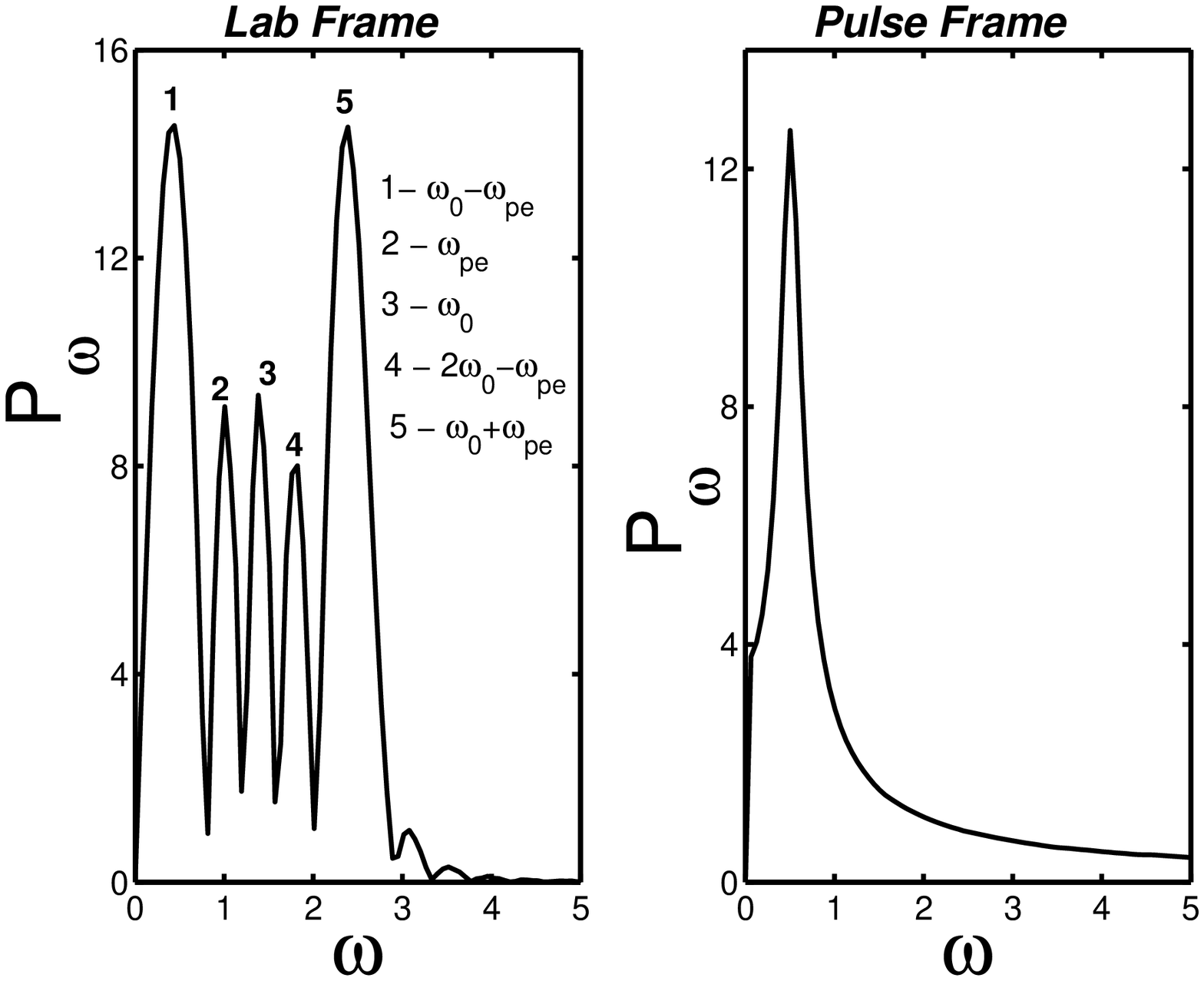}
\end{figure}

\newpage
\begin{figure}
\includegraphics[]{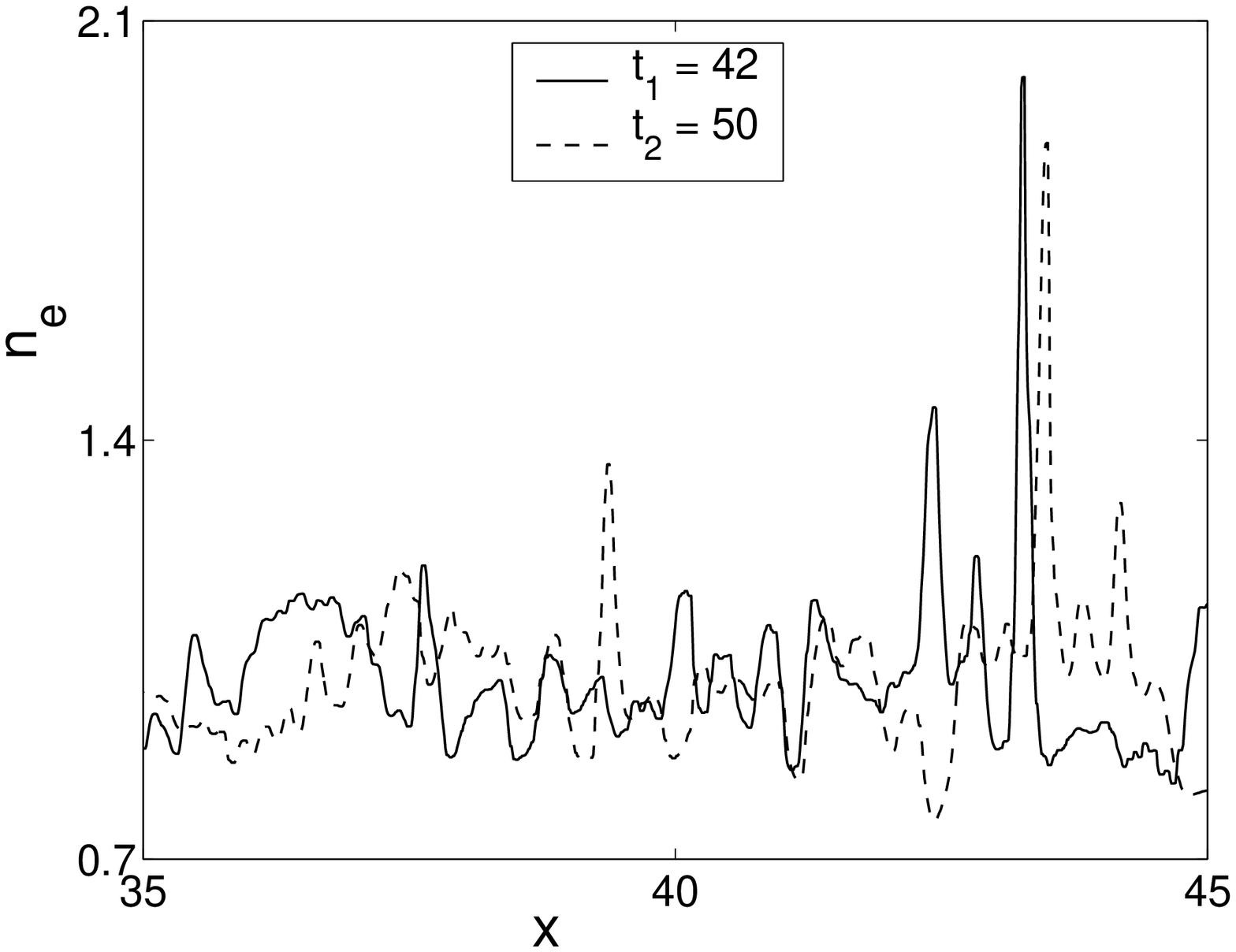}
\end{figure}

\newpage
\begin{figure}
\includegraphics[]{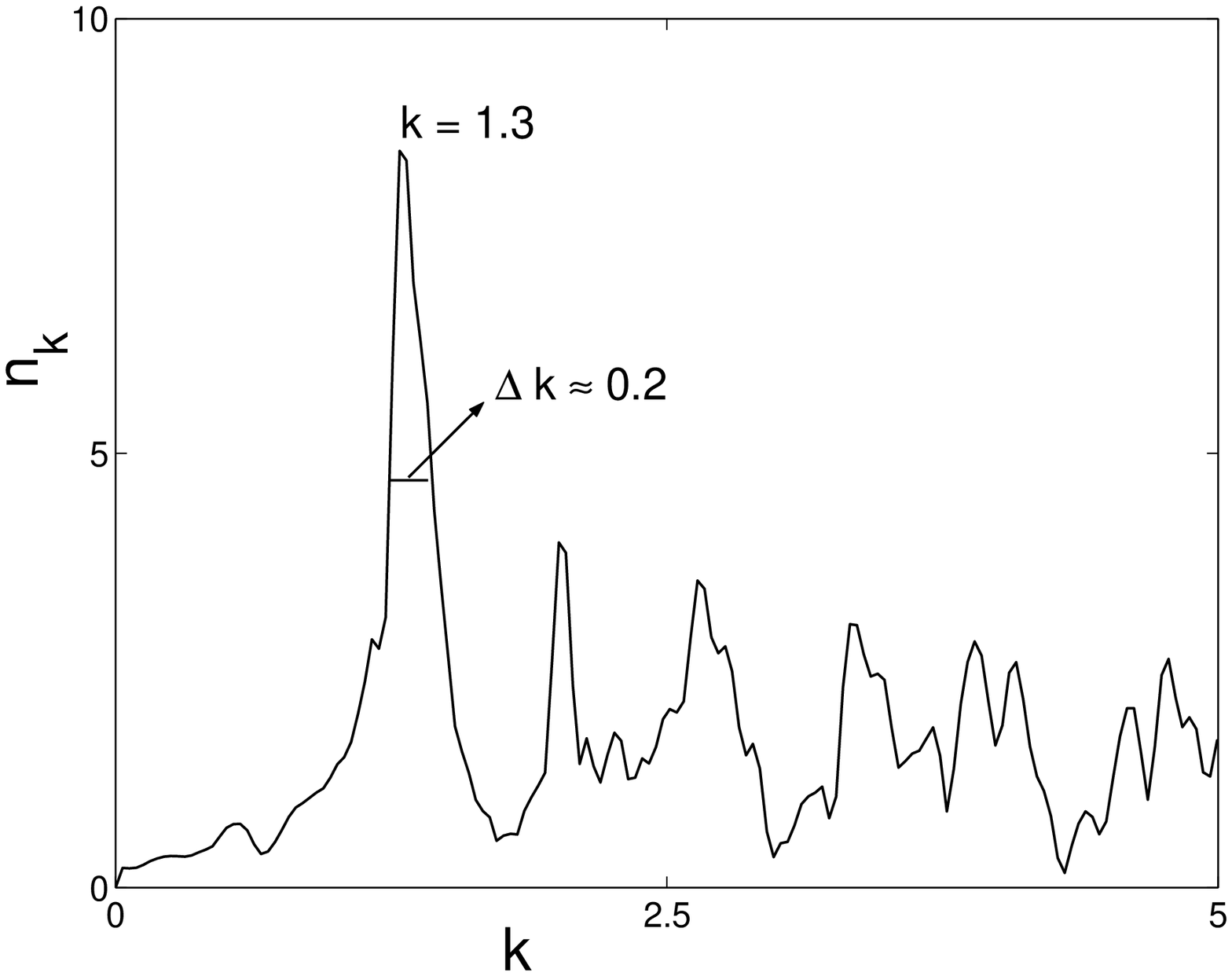}
\end{figure}

\end{document}